\documentstyle[floats,prl,aps]{revtex}
\begin{document}
\renewcommand{\thefootnote}{\fnsymbol{footnote}}
\draft
\title{\large\bf 
Integrable eight-state supersymmetric $U$ model with boundary terms and
    its Bethe ansatz solution} 

\author{ Xiang-Yu Ge \footnote{E-mail:xg@maths.uq.edu.au},
	Mark D. Gould, Yao-Zhong Zhang \footnote {Queen Elizabeth II Fellow.
                      E-mail: yzz@maths.uq.edu.au}
             and 
        Huan-Qiang Zhou \footnote{On leave of absence from Dept of
	         Physics, Chongqing University, Chongqing 630044, China.
                 E-mail: hqzhou@cqu.edu.cn}} 

\address{Department of Mathematics,University of Queensland,
		     Brisbane, Qld 4072, Australia}

\maketitle

\vspace{10pt}

\begin{abstract}
A class of integrable boundary terms for the eight-state supersymmtric
$U$ model are presented by solving the graded reflection equations.
The boundary model is solved by using the coordinate Bethe ansatz method
and the Bethe ansatz equations are obtained. 
\end{abstract}




\def\a{\alpha}
\def\b{\beta}
\def\d{\delta}
\def\e{\epsilon}
\def\g{\gamma}
\def\k{\kappa}
\def\l{\lambda}
\def\o{\omega}
\def\t{\theta}
\def\s{\sigma}
\def\D{\Delta}
\def\L{\Lambda}


\def\beq{\begin{equation}}
\def\eeq{\end{equation}}
\def\bea{\begin{eqnarray}}
\def\eea{\end{eqnarray}}
\def\ba{\begin{array}}
\def\ea{\end{array}}
\def\no{\nonumber}
\def\le{\langle}
\def\re{\rangle}
\def\lt{\left}
\def\rt{\right}

\newcommand{\reff}[1]{eq.~(\ref{#1})}

\vskip.3in
Integrable systems with boundary interactions are one of the recent
achievements which the authors think deserve careful investigations.
Here we only concern with 1-dimensional open-boundary lattice integrable
models of strongly correlated electrons. Such lattice models can be
treated by Sklyanin's boundary quantum inverse scattering method (QISM) 
\cite{Skl88} or its generalizations \cite{Mez91,deV93,Zho96,Bra97}. 
More specifically, we present integrable boundary terms for the
eight-state version of the supersymmetric
$U$ model recently introduced in \cite{Gou97}. 
The bulk Hamiltonian describes a supersymmetric electron model with
correlated single-particle
and pair hoppings as well as uncorrelated triple-particle hopping. 
So the model on an open lattice, which we consider here, involves
many physically interesting processes with boundary
interactions. The boundary model is solved by means of the coordinate Bethe 
ansatz method and the Bethe ansatz equations are derived.

Let $c_{j,\a}^\dagger$ ($c_{j,\a}$) denote fermionic creation 
(annihilation) operator which creates (annihilates) an electron of
species $\a ~(=+,0,-)$ at
site $j$. They satisfy the anti-commutation relations given by
$\{c_{i,\a}^\dagger, c_{j,\b}\}=\d_{ij}\d_{\a\b}$, where 
$i,j=1,2,\cdots,L$ and $\a,\b=+,\;0,\;-$. We consider the
following Hamiltonian with boundary terms
\beq
H=\sum _{j=1}^{L-1} H_{j,j+1}^{\rm bulk} + H^{\rm boundary}_L 
+H^{\rm boundary}_R,\label{h}
\eeq
where $H^{\rm bulk}_{j,j+1}$ is the Hamiltonian density of the eight-state 
supersymmetric $U$ model \cite{Gou97}
\bea
H^{\rm bulk}_{j,j+1}(g)&=&-\sum_\a(c_{j,\a}^\dagger c_{j+1,\a}+{\rm h.c.})
  \;\exp\lt\{-\frac{\eta}{2} \sum_{\b(\neq\a)}(n_{j,\b}+n_{j+1,\b})
  +\frac{\zeta}{2}
  \sum_{\b\neq\g(\neq\a)}(n_{j,\b}n_{j,\g}
  +n_{j+1,\b} n_{j+1,\g})\rt\}\no\\
& &-\frac{1}{2(g+1)}\sum_{\a\neq\b\neq\g}(c_{j,\a}^\dagger c_{j,\b}^\dagger 
  c_{j+1,\b}c_{j+1,\a}+{\rm h.c.})
  \;\exp\lt\{-\frac{\xi}{2}  (n_{j,\g}+n_{j+1,\g})\rt\}\no\\
& &-\frac{2}{(g+1)(g+2)}\lt(c_{j,+}^\dagger c_{j,0}^\dagger 
  c_{j,-}^\dagger c_{j+1,-} c_{j+1,0} c_{j+1,+}+{\rm h.c.}\rt)\no\\
& & +\sum_\a (n_{j,\a}+n_{j+1,\a})-\frac{1}{2(g+1)}\sum_{\a\neq\b}
  (n_{j,\a}n_{j,\b}+n_{j+1,\a}n_{j+1,\b})\no\\
& &+\frac{2}{(g+1)(g+2)}(n_{j,+}n_{j,0}n_{j,-}+n_{j+1,+}n_{j+1,0}
  n_{j+1,-})\label{hamiltonian}
\eea
with
\beq
\eta=-\ln\frac{g}{g+1},~~~\zeta=\ln(g+1)-\frac{1}{2}
  \ln g(g+2),~~~\xi=-\ln\frac{g}{g+2},
\eeq
and $H^{\rm boundary}_L,~H^{\rm boundary}_R$ are boundary terms
\bea
H^{\rm boundary}_L&=&-\frac {2g}
{2-\xi_-}\lt(n_1-
\frac {2}{\xi_-}
 ( n_{1+}n_{10}+n_{10}n_{1-}+n_{1+}n_{1-})
  +\frac {8}{\xi _- (2+\xi _-)}n_{1+} n_{10} n_{1-} \rt),\no\\
H^{\rm boundary}_R&=&
-\frac {2g}{2-\xi_+}\lt(n_L-
\frac {2}{\xi_+}
 ( n_{L+}n_{L0}+n_{L0}n_{L-}+n_{L+}n_{L-})
  +\frac {8}{\xi _+ (2+\xi _+)}n_{L+} n_{L0} n_{L-} \rt)
  .\label{bounadry-terms}
\eea
In the above equations, $n_{j\a}$ is the number density operator
$n_{j\a}=c_{j\a}^{\dagger}c_{j\a}$,
$n_j=n_{j+}+n_{j0}+n_{j-}$; 
$\xi_{\pm}$ are some parameters describing the
boundary effects.

As was shown in \cite{Gou97}, the supersymmetry algebra underlying the bulk
model is $gl(3|1)$. The boundary terms may spoil this $gl(3|1)$ supersymmetry,
leaving the boundary model with a smaller symmetry algebra. If one 
projects out any one species, then the projected Hamiltonian is 
nothing but that of the supersymmetric $U$ model with boundary terms
\cite{Zha97} with the following identification of the parameters
in the two models: $U=\pm\frac{2}{g+1}$.

We will establish the quantum integrability of the Hamiltonian for
the boundary eight-state
supersymmetric $U$ model (\ref{h}) by showing that it can be derived from
the (graded) boundary quantum inverse scattering method. 
Let us first of all recall that the
Hamiltonian of 
the eight-state supersymmetric $U$ model with the periodic boundary conditions
commutes with the bulk transfer matrix, which is the supertrace of the
monodromy matrix $T(u)$,
\beq
T(u) = R_{0L}(u)\cdots R_{01}(u). \label{matrix-t}
\eeq
where the subscript $0$ denotes the auxiliary superspace $V=C^{4,4}$.
It should be noted that the supertrace
is carried out for the auxiliary superspace $V$. The R-matrix
$ R(u)\equiv P\check{R}(u)$, where $P$ is graded permutation operator,
is given by 
\beq
\check{R}(u)=\check{P}_1-\frac{u+2g}{u-2g}\check{P}_2
  +\frac{(u+2g)(u+2g+2)}{(u-2g)(u-2g-2)}\check{P}_3
  -\frac{(u+2g)(u+2g+2)(u+2g+4)}{(u-2g)(u-2g-2)(u-2g-4)}\check{P}_4,
  \label{rational-R}
\eeq
where $\check{P}_k,~k=1,2,3,4$, are four projection operators whose
explicit formulae can be found in \cite{Gou97}.
The elements of the supermatrix $T(u)$ are the generators
of an associative superalgebra ${\cal A}$ defined by the relations
\beq
R_{12}(u_1-u_2) \stackrel {1}{T}(u_1) \stackrel {2}{T}(u_2) =
   \stackrel {2}{T}(u_2) \stackrel {1}{T}(u_1)R_{12}(u_1-u_2),\label{rtt-ttr} 
\eeq
where $\stackrel {1}{X} \equiv  X \otimes 1,~
\stackrel {2}{X} \equiv  1 \otimes X$
for any supermatrix $ X \in End(V) $. For later use, we list some useful
properties enjoyed by the R-matrix:
(i) Unitarity:   $  R_{12}(u)R_{21}(-u) = 1$ and (ii)
 Crossing-unitarity:  $  R^{st_2}_{12}(-u+4)R^{st_1}_{21}(u) =
         \tilde {\rho }(u)$
with $\tilde \rho (u)$ being a scalar function.

In order to construct integrable electronic models with open boundary
conditions, we introduce the following graded reflection equations
(REs) that the so-called boundary
K-(super)matrices $K_\pm(u)$ satisfy
\beq
R_{12}(u_1-u_2)\stackrel {1}{K}_-(u_1) R_{21}(u_1+u_2)
  \stackrel {2}{K}_-(u_2)
=  \stackrel {2}{K}_-(u_2) R_{12}(u_1+u_2)
  \stackrel {1}{K}_-(u_1) R_{21}(u_1-u_2),  \label{reflection1}
\eeq
\bea
&&R_{21}^{st_1 ist_2}(-u_1+u_2)\stackrel {1}{K_+^{st_1}}
  (u_1) R_{12}(-u_1-u_2+4)
  \stackrel {2}{K_+^{ist_2}}(u_2)\no\\
&&~~~~~~~~~~~~~~~~~~~~~=\stackrel {2}{K_+^{ist_2}}(u_2) R_{21}(-u_1-u_2+4)
  \stackrel {1}{K_+^{st_1}}(u_1) R_{12}^{st_1 ist_2}(-u_1+u_2)
  ,\label{reflection2}
\eea
where the supertransposition $st_{\mu}~(\mu =1,2)$ 
is only carried out in the
$\mu$-th factor superspace of $V \otimes V$, whereas $ist_{\mu}$ denotes
the inverse operation of  $st_{\mu}$. 
Following  Sklyanin's arguments \cite{Skl88}, one
may show that the quantity ${\cal T}_-(u)$ given by
\beq
{\cal T}_-(u) = T(u) {K}_-(u) T^{-1}(-u) 
\eeq
satisfies the same relation as $K_-(u)$:
\beq
R_{12}(u_1-u_2)\stackrel {1}{\cal T}_-(u_1) R_{21}(u_1+u_2)
  \stackrel {2}{\cal T}_-(u_2)
=  \stackrel {2}{\cal T}_-(u_2) R_{12}(u_1+u_2)
  \stackrel {1}{\cal T}_-(u_1) R_{21}(u_1-u_2).
\eeq
Thus if one defines the boundary transfer matrix $\tau(u)$ as
\beq
\tau(u) = str (K_+(u){\cal T}_-(u))=str\lt(K_+(u)T(u)K_-(u)T^{-1}(-u)\rt),
\eeq
then it can be shown \cite{Bra97} that
\beq
[\tau(u_1),\tau(u_2)] = 0.
\eeq

We now solve (\ref{reflection1}) and (\ref{reflection2}) 
for $K_+(u)$ and $K_-(u)$.
Let us restrict ourselves to the diagonal solutions. Then, one may 
check that the matrix $K_-(u)$ given by
\beq
K_-(u)=  \frac {1}{ \xi_-(2-\xi_-)(2+\xi_-)} \left ( \begin {array}
{cccccccc}
A_-(u)&0&0&0&0&0&0&0\\
0&B_-(u)&0&0&0&0&0&0\\
0&0&B_-(u)&0&0&0&0&0\\
0&0&0&B_-(u)&0&0&0&0\\
0&0&0&0&C_-(u)&0&0&0\\
0&0&0&0&0&C_-(u)&0&0\\
0&0&0&0&0&0&C_-(u)&0\\
0&0&0&0&0&0&0&D_-(u)
\end {array} \right ),\label{k-}
\eeq
where
\bea
A_-(u)&=&(-\xi_-+u)(2-\xi_-+u)(-2-\xi_- +u),\no\\
B_-(u)&=&(-\xi_-+u)(2-\xi_--u)(-2-\xi_- +u),\no\\
C_-(u)&=&(-\xi_--u)(2-\xi_--u)(-2-\xi_-+u),\no\\
D_-(u)&=&(-\xi_--u)(2-\xi_--u)(-2-\xi_--u),
\eea
satisfies (\ref{reflection1}).
The matrix $K_+(u)$ can be obtained from the isomorphism of the
two REs. Indeed, given a solution
$K_- (u)$ of (\ref{reflection1}), then $K_+(u)$ defined by
\beq
K_+^{st}(u) =  K_-(-u+2)\label{t+t-}
\eeq
is a solution of (\ref{reflection2}). 
The proof follows from some algebraic computations upon
substituting (\ref{t+t-}) into  
(\ref{reflection2}) and making use
of the properties of the R-matrix .
Therefore, one may choose the boundary matrix $K_+(u)$ as 
\beq
K_+(u)=   \left ( \begin {array}
{cccccccc}
A_+(u)&0&0&0&0&0&0&0\\
0&B_+(u)&0&0&0&0&0&0\\
0&0&B_+(u)&0&0&0&0&0\\
0&0&0&B_+(u)&0&0&0&0\\
0&0&0&0&C_+(u)&0&0&0\\
0&0&0&0&0&C_+(u)&0&0\\
0&0&0&0&0&0&C_+(u)&0\\
0&0&0&0&0&0&0&D_+(u)
\end {array} \right ),\label{k+}
\eeq
where
\bea
A_+(u)&=&(-2g +2 +\xi_+-u)(-2g  +\xi_+-u)(-2g-2+\xi _+ -u),\no\\
B_+(u)&=&(-2g -2 +\xi_++u)(-2g  +\xi_+-u)(-2g-2+\xi _+ -u),\no\\
C_+(u)&=&(-2g -2 +\xi_++u)(-2g -4 +\xi_++u)(-2g-2+\xi _+ -u),\no\\
D_+(u)&=&(-2g -2 +\xi_++u)(-2g  -4+\xi_++u)(-2g-6+\xi _+ +u).
\eea

Now it can be shown  that 
Hamiltonian (\ref{h}) is related to the second derivative of
the boundary transfer matrix
$\tau (u)$ (up to an unimportant additive constant)
\bea
H&=&2g \; H^R,\no\\
H^R&=&\frac {\tau'' (0)}{4(V+2W)}=
  \sum _{j=1}^{L-1} H^R_{j,j+1} + \frac {1}{2} \stackrel {1}{K'}_-(0)
+\frac {1}{2(V+2W)}\lt[str_0\lt(\stackrel {0}{K}_+(0)G_{L0}\rt)\rt.\no\\
& &\lt.+2\,str_0\lt(\stackrel {0}{K'}_+(0)H_{L0}^R\rt)+
  str_0\lt(\stackrel {0}{K}_+(0)\lt(H^R_{L0}\rt)^2\rt)\rt],\label{derived-h}
\eea
where 
\bea
V&=&str_0 K'_+(0),
~~~~~W=str_0 \lt(\stackrel {0}{K}_+(0) H_{L0}^R\rt),\no\\
H^R_{i,j}&=&P_{i,j}R'_{i,j}(0),
~~~~~G_{i,j}=P_{i,j}R''_{i,j}(0).
\eea
Here $P_{i,j}$ denotes the graded permutation operator acting on the $i$-th
and $j$-th quantum spaces. (\ref{derived-h}) implies that the boundary
eight-state supersymmetric $U$ model admits
an infinite number
of conserved currents which are in involution with each other, thus
assuring its integrability. It should be emphasized that 
Hamiltonian (\ref{h}) appears as the second derivative of the boundary
transfer matrix
$\tau (u)$ with respect to the spectral parameter $u$ at $u=0$. This
is due to the fact that the supertrace of $K_+(0)$ equals to zero.
The reason for the zero supertrace of $K_+(0)$
is related to the fact that the quantum space is the
8-dimensional {\em typical} irreducible representation of $gl(3|1)$. 
A similar situation has already appeared in many spin chain
models \cite{Zho96,Zha97,Bra97}.

Having established the quantum integrability of the boundary model,
we now  solve it by using
the coordinate space Bethe ansatz method. The whole procedure is
similar to that for other models \cite{Asa96,Zha97}.
The Bethe ansatz equations are 
\bea
e^{ik_j2(L+1)}\zeta(k_j;p_1)\zeta(k_j;p_L)
&=&\prod_{\s=1}^{M_1}\frac{\t_j-\l^{(1)}_\s+ic/2}
      {\t_j-\l^{(1)}_\s-ic/2}
\frac{\t_j+\l^{(1)}_\s+ic/2}
      {\t_j+\l^{(1)}_\s-ic/2},\no\\
\prod_{j=1}^N\frac{\l^{(1)}_\s-\t_j+ic/2}{\l^{(1)}_\s-\t_j-ic/2}
\frac{\l^{(1)}_\s+\t_j+ic/2}{\l^{(1)}_\s+\t_j-ic/2}&=&
   -\prod_{\rho=1}^{M_1}\frac{\l^{(1)}_\s-\l^{(1)}_\rho+ic}
   {\l^{(1)}_\s-\l^{(1)}_\rho-ic}
   \frac{\l^{(1)}_\s+\l^{(1)}_\rho+ic}
   {\l^{(1)}_\s+\l^{(1)}_\rho-ic}
   \prod_{\rho=1}^{M_2}\frac{\l^{(1)}_\s-\l^{(2)}_\rho-ic/2}
   {\l^{(1)}_\s-\l^{(2)}_\rho+ic/2}
   \frac{\l^{(1)}_\s+\l^{(2)}_\rho-ic/2}
   {\l^{(1)}_\s+\l^{(2)}_\rho+ic/2},\no\\
   & &\s=1,\cdots,M_1,\no\\
\prod_{\rho=1}^{M_1}\frac{\l^{(2)}_\g-\l^{(1)}_\rho+ic/2}
   {\l^{(2)}_\g-\l^{(1)}_\rho-ic/2}
\frac{\l^{(2)}_\g+\l^{(1)}_\rho+ic/2}
   {\l^{(2)}_\g+\l^{(1)}_\rho-ic/2}&=&
   \prod_{\rho=1}^{M_2}\frac{\l^{(2)}_\g-\l^{(2)}_\rho+ic}
   {\l^{(2)}_\g-\l^{(2)}_\rho-ic}
   \frac{\l^{(2)}_\g+\l^{(2)}_\rho+ic}
   {\l^{(2)}_\g+\l^{(2)}_\rho-ic},\no\\
  & & \g=1,\cdots,M_2,\label{Bethe-ansatz}
\eea
where 
\beq
\zeta (k;p)= \frac{1-pe^{-ik}}{1-pe^{ik}},~~~~
  p_1=-1-\frac {2g}{2-\xi_-},~~~~
  p_L=-1-\frac {2g}{2-\xi_+},
\eeq
and $c=e^\eta-1$;
the charge rapidities  $\t_j\equiv \t(k_j)$ 
are related to the single-particle
quasi-momenta $k_j$ by $\theta (k)=\frac {1}{2} \tan (\frac {k}{2})$.
The energy eigenvalue $E$ of the model is given by
$E=-2\sum ^N_{j=1}\cos k_j$ (modular an unimportant additive constant coming
from the chemical potential term).

In conclusion, we have studied integrable open-boundary conditions for the
eight-state supersymmetric $U$ model. The  quantum integrability of the
system follows from the fact
that the Hamiltonian may be embedded into
a one-parameter family of commuting transfer matrices. Moreover, the Bethe
ansatz equations are derived by use of the coordinate space Bethe ansatz
approach. This provides us with a basis for computing the finite size
corrections (see, e.g. \cite{Asa96,Yun95})
to the low-lying energies in the system, which in turn allow
us to use the boundary conformal field theory technique to study
the critical properties of the boundary.
The details will be treated in a separate publication.

\vskip.3in
This work is supported by Australian Research Council, University of
Queensland New Staff Research Grant and Enabling Research Grant. H.-Q.Z
would like to thank Department of Mathematics of UQ for kind hospitality. He
also thanks the supports from the National Natural Science Foundation
of China and from Sichuan Young Investigators Science and Technology
Fund.


\end{document}